\documentclass[a4paper,11pt]{article}

\usepackage{hyperref}
\usepackage{graphicx}
\usepackage{amssymb}
\usepackage{amsfonts}
\usepackage{amsbsy}
\usepackage{amsmath}
\usepackage{longtable}
\usepackage{a4wide}
\usepackage{color}

\begin{document}
\thispagestyle{empty}
\begin{flushright}
IPMU11-0044\\
TUW-11-03
\end{flushright}
\vspace{1cm}
\begin{center}
{\bf\Large Toric Methods in F-theory Model Building}
\end{center}
\vspace{8mm}
\begin{center}
{\large Johanna Knapp${}^{\dagger\ast}$\footnote{e-mail: johanna.knapp AT ipmu.jp}, Maximilian Kreuzer${}^{\ast}$}
\end{center}
\vspace{3mm}
\begin{center}
{\it $^{\dagger}$ Institute for the Physics and Mathematics of the Universe (IPMU)\\
The University of Tokyo \\
5-1-5 Kashiwanoha\\
Kashiwa 277-8583\\
Japan
}
\end{center}
\begin{center}
{\it $^{\ast}$Institut f\"ur theoretische Physik \\
TU Vienna\\
Wiedner Hauptstrasse 8-10\\
1040 Vienna\\
Austria }
\end{center}
\begin{center}
\end{center}
\vspace{15mm}
\begin{abstract}
\noindent In this review article we discuss recent constructions of global F-theory GUT models and explain how to make use of toric geometry to do calculations within this framework. After introducing the basic properties of global F-theory GUTs we give a self-contained review of toric geometry and introduce all the tools that are necessary to construct and analyze global F-theory models. We will explain how to systematically obtain a large class of compact Calabi-Yau fourfolds which can support F-theory GUTs by using the software package PALP.
\end{abstract}
\newpage
\setcounter{tocdepth}{1}
\tableofcontents
\setcounter{footnote}{0}
\section{Introduction}
Even though it has been around for quite a while \cite{Vafa:1996xn}, F-theory has recently received a lot of new attention as a setup where Grand Unified Theories (GUTs) can be conceived from string theory. Starting with \cite{Donagi:2008ca,Beasley:2008dc,Beasley:2008kw} the phenomenology of F-theory GUTs has become an active field of research. The basic idea is that the GUT theory is localized on a $(p,q)$ seven-brane $S$ inside a three-dimensional base manifold $B$ of an F-theory compactification on an elliptically fibered Calabi-Yau fourfold. The location of the GUT brane and the gauge group are determined by the degeneration of the elliptic fibration. Chiral matter localizes on curves inside the GUT brane $S$ where gauge enhancement occurs, Yukawa couplings sit at points. For many phenomenological applications it is sufficient to consider the field theory living on the GUT brane without specifying the details of the global F-theory compactification. However, fluxes, monodromies or consistency constraints such as tadpole cancellation cannot be addressed in a purely local setup. These issues have recently received a lot of attention in the literature \cite{Andreas:2009uf,Blumenhagen:2009yv,Marsano:2009wr,Grimm:2009yu,Blumenhagen:2010ja,Cvetic:2010rq,Hayashi:2010zp,Chen:2010tp,Chen:2010ts,Grimm:2010ez,Marsano:2010ix,Chung:2010bn,Heckman:2010qv,Cvetic:2010ky,Cecotti:2010bp,Marsano:2010sq,Collinucci:2010gz,Chiou:2011js,Ludeling:2011en,Knapp:2011wk,Dolan:2011iu}. Therefore it is interesting to see whether it is possible to embed the local F-theory GUT into a compactification on a Calabi-Yau fourfold. Most known examples of compact Calabi-Yau manifolds are hypersurfaces or complete intersections in a toric ambient space. It is thus natural to look for Calabi-Yau fourfolds within this class of examples. A prescription for constructing elliptically fibered Calabi-Yau fourfolds as complete intersections in a six-dimensional toric ambient space has been given in \cite{Blumenhagen:2009yv,Grimm:2009yu}. Before that complete intersection Calabi-Yau fourfolds in F-theory had already been used in the context of F-theory uplifts of type IIB string theory \cite{Collinucci:2008zs,Collinucci:2009uh,Blumenhagen:2009up}. A similar construction has also been discussed in \cite{Cvetic:2010rq}. It has been shown in examples that it is indeed possible to construct viable F-theory GUTs within this framework. 

The construction of \cite{Blumenhagen:2009yv} is very well-suited for a systematic search of a large class of models. This is interesting for several reasons: one goal is to find particularly nice examples of F-theory compactifications. Even though the known examples have been able to incorporate F-theory models, one usually gets much more than just that. In minimal F-theory GUTs one typically needs only very few Yukawa points and a small number of moduli on the matter curves. This is not satisfied in most known global models. A related question deals with the genericity of F-theory GUTs. The geometric configurations used for constructing such models are usually quite special and one may wonder how often they can be realized in elliptically fibered fourfolds. From the point of view of model building it is useful to have some easy-to-check geometric conditions which makes it possible to select suitable models from a large class of geometries. This will be discussed in more details in the text. From a mathematical point of view it might be interesting to obtain a partial classification of Calabi-Yau fourfolds.

This review article discusses selected topics in toric geometry and F-theory GUTs. The paper is organized as follows: in section \ref{sec-fth} we recall the construction of global F-theory models and discuss the basic requirements we would like to impose. In section \ref{sec-toric} we review several notions in toric geometry which are required in order to perform the F-theory calculations. The geometries one has to deal with are usually quite complicated, and very often one has to rely on computer support in order to be able to do explicit calculations. Therefore we discuss how such calculations can be implemented using existing software such as PALP \cite{Kreuzer:2002uu}. We will mainly focus on the application of toric geometry in the context of F-theory model building. For a more complete picture on this vast subject of F-theory phenomenology we refer to other review articles such as \cite{Denef:2008wq,Heckman:2010bq,Weigand:2010wm}. For more extensive discussions of toric geometry we recommend \cite{fultontoric,Kreuzer:2006ax,coxbook}.
\section{Global F-theory models}
\label{sec-fth}
\subsection{Setup}
In this section we introduce the basic concepts and notions used in global F-theory models. In the remainder of this review we will explain the techniques which are necessary to do calculations within this framework. For more details on how the quantities introduced below come about we refer to the original papers or the recent review article \cite{Weigand:2010wm}.

In \cite{Blumenhagen:2009yv} it has been proposed to construct Calabi-Yau fourfolds, which are suitable for F-Theory model building, as complete intersections of two hypersurfaces in a six-dimensional toric ambient space. The hypersurface equations have the following structure:
\begin{equation}
\label{cicydef}
P_{B}(y_i,w)=0\qquad\qquad P_W(x,y,z,y_i,w)=0
\end{equation}
The first equation only depends on the homogeneous coordinates $(y_i,w)$ of the three-dimensional base $B$ of the elliptically fibered Calabi-Yau fourfold $X_4$. Here we have singled out one coordinate $w$, indicating that the divisor given by $w=0$ defines a seven-brane $S$ which supports a GUT theory of the type introduced in \cite{Donagi:2008ca,Beasley:2008dc,Beasley:2008kw}. The second equation in (\ref{cicydef}) defines a Weierstrass model, where $(x,y,z)$ are those coordinates of the six-dimensional ambient space that describe the torus fiber. For this type of elliptic fibrations $P_W$ is of Tate form which is defined as follows:
\begin{equation}
\label{tate}
P_W=x^3-y^2+xyz a_1+x^2z^2a_2+yz^3a_3+xz^4a_4+z^6a_6
\end{equation}
The $a_n(y_i,w)$ are sections of $K_B^{-n}$, where $K_B$ is the canonical bundle of the base manifold. Furthermore $x$ and $y$ can be viewed as sections of $K_B^{-2}$ and $K_B^{-3}$, respectively.

The information about the F-theory model is encoded in the Tate equation (\ref{tate}). In order to have a non-trivial gauge group on the GUT brane the elliptic fibration must degenerate above $S$. The gauge group is determined by the structure of the singularity. The elliptic fibration becomes singular at the zero locus of the discriminant $\Delta$. Defining the polynomials $\beta_2=a_1^2+4a_2$, $\beta_4=a_1a_3+2a_4$ and $\beta_6=a_3^2+4a_6$ the discriminant is given by the following expression:
\begin{equation}
\Delta=-\frac{1}{4}\beta_2^2(\beta_2\beta_6-\beta_4^2)-8\beta_4^3-27\beta_6^2+9\beta_2\beta_4\beta_6
\end{equation}
 According to Kodaira's classification \cite{Kodaira} and Tate's algorithm \cite{tate}, the gauge group can be inferred from the factorization properties of the $a_n(y_i,w)$ with respect to $w$. Considering for instance $SU(5)$- and $SO(10)$-models, the factorization looks like this:
\begin{eqnarray}
\label{su5so10tate}
SU(5):&&a_1=b_5w^0\quad a_2=b_4w^1\quad a_3=b_3w^2\quad a_4=b_2w^3\quad a_6=b_0w^5\nonumber\\
SO(10):&& a_1=b_5w^1\quad a_2=b_4w^1\quad a_3=b_3w^2\quad a_4=b_2w^3\quad a_6=b_0w^5
\end{eqnarray}
The $b_i$s are sections of some appropriate line bundle over $B$ that have at least one term independent of $w$. 

In F-theory GUT models chiral matter localizes on curves on $S$, where a rank $1$ enhancement of the gauge group appears. In $SU(5)$ models the matter curves are at the following loci inside $S$:
\begin{equation}
b_3^2b_4-b_2b_3b_5+b_0b_5^2=0\quad \textrm{{\bf 5} matter} \qquad b_5=0\quad \textrm{{\bf 10} matter}\,,
\end{equation}
The matter curves for the $SO(10)$ models are at:
\begin{equation}
b_3=0\quad \textrm{{\bf 10} matter} \qquad b_4=0\quad \textrm{{\bf 16} matter}\,.
\end{equation}
Yukawa coupling arise at points inside $S$ where the GUT singularity has an rank $2$ enhancement. In $SU(5)$ models the Yukawa points sit at:
\begin{eqnarray}
 b_4=0 \cap b_5=0&\quad& {\bf 10\:10\:5} \textrm{ Yukawas}\quad \textrm{$E_6$ enhancement} \nonumber\\ 
b_5=0 \cap b_3=0&\quad& {\bf 10\:\bar{5}\:\bar{5}} \textrm{ Yukawas} \quad \textrm{$SO(12)$ enhancement}\,.
\end{eqnarray}
In the $SO(10)$-case we have the following Yukawa couplings:
\begin{eqnarray}
b_3=0 \cap b_4=0&\quad& {\bf 16\:16\:10} \textrm{ Yukawas}\quad \textrm{$E_7$ enhancement} \nonumber\\ 
b_2^2-4 b_0 b_4=0 \cap b_3=0&\quad& {\bf 16\:10\:10} \textrm{ Yukawas} \quad \textrm{$SO(14)$ enhancement}\,.
\end{eqnarray}
Given a complete intersection Calabi-Yau fourfold of the form (\ref{cicydef}) the expressions for matter curves and Yukawa points are globally defined and can be calculated explicitly. Having a global F-theory compactification we can furthermore calculate the Hodge numbers and the Euler number $\chi_4$ of the Calabi-Yau fourfold $X_4$. The latter enters the D3-tadpole cancellation condition, 
\begin{equation}
\frac{\chi_4}{24}=N_{D3}+\frac{1}{2}\int_{X_4}G_4\wedge G_4,
\end{equation} 
where $G_4$ denotes the fourform flux on $X_4$ and $N_{D3}$ is the number of $D3$-branes.
\subsection{Geometric Data in F-theory models}
\label{sec-fgeom}
So far we have summarized the basic structure of a global F-theory GUT. In the present section we will discuss which properties of the GUT model are encoded in the geometries of the base manifold $B$ and the Calabi-Yau fourfold $X_4$. We will not go deeply into the phenomenology of F-theory GUTs but rather focus on the basic geometric properties which should be satisfied in order to obtain a viable GUT model. 
\subsubsection*{Base Manifold}
Since the GUT brane $S$ is a divisor in a three-dimensional base manifold $B$, a large amount of information about the model can be extracted from the geometry of $B$. The base $B$ is a non-negatively curved manifold of complex dimension three. In our setup it will be given by a hypersurface in a toric ambient space. Note that Fano threefolds are not a good choice for $B$ due to the lack of a decoupling limit \cite{Cordova:2009fg}. In section \ref{sec-toric} we discuss a systematic construction of such base manifolds using toric geometry. In order to have a well defined model we have to make sure that $B$ is non-singular. In contrast to Calabi-Yau threefolds the base manifolds for F-theory GUTs may inherit the singularities of the ambient space. Therefore checks for the regularity of $B$ have to be implemented.

Having found a suitable base manifold the next step is to identify divisors inside $B$ that can support GUT models. The most promising candidates for F-theory model building are del Pezzo surfaces. These are Fano twofolds (see for instance \cite{Griffiths:1994ab}).  Note however, that del Pezzos are not the only possibility for the construction of GUT models in F-theory. See \cite{Braun:2010hr} for a recent discussion. There are several motivations to focus on del Pezzo divisors. In local F-theory GUTs the del Pezzo property ensures the existence of a decoupling limit \cite{Beasley:2008dc,Beasley:2008kw}. For $SU(5)$ GUT models, the fact that del Pezzos have $h^{0,1}=h^{2,0}=0$ implies some powerful vanishing theorems which forbid exotic matter after breaking $SU(5)$ to the Standard Model gauge group \cite{Beasley:2008kw}. 

We can identify candidates for del Pezzo divisors inside $B$ by their topological data.  Suppose the base manifold $B$ is embedded in a toric ambient space which has toric divisors $D_i$. The $D_i$ give a homology basis of the ambient space. In this setup the hypersurface is specified by a divisor, which we will by abuse of notation also call $B$, that is given in terms of a linear combination of the $D_i$.  The total Chern class of a particular divisor $S$ in the ambient space is, after restriction to $B$ (for more details see section \ref{sec-intmori}):
\begin{equation}
c(S)=\frac{\prod_i(1+D_i)}{(1+B)(1+S)}
\end{equation}
In order to apply this formula we have to know the intersection ring of $B$. As we will discuss in section \ref{sec-intmori} this can be obtained from the intersection ring of the ambient space.

A necessary condition for a divisor $S$ to be $dP_n$ is that it must have the following topological data:
\begin{equation}
\int_Sc_1(S)^2=9-n\qquad \int_Sc_2(S)=n+3\qquad\Rightarrow\qquad \chi_h=\int_S\mathrm{Td}(S)=1,
\end{equation}
where $\chi_h=\sum_i(-1)^ih^{0,i}$ is the holomorphic Euler characteristic and $\mathrm{Td}$ denotes the Todd class. In the equations above the integration over the four-cycle (representing the divisor) $S$ is equivalent to computing the intersection with $S$.
Since del Pezzos are Fano twofolds, we have a further necessary condition: the intersections of $c_1(S)$ with curves on $S$ have to be positive. In the toric setup we can only check this for curves which are induced from the divisors on the ambient space. In that case the condition is:
\begin{equation}
\label{poscurve}
D_i\cdot S\cdot c_1(S) > 0\qquad D_i\neq S \qquad \forall D_i\cdot S\neq\emptyset\,.
\end{equation}
In order to make these calculations we need to know the homology basis of toric divisors and their intersection numbers. 

In local F-theory GUTs the del Pezzo property is sufficient to ensure the existence of a decoupling limit. For global models further checks are in order. Gravity decouples from the gauge degrees of freedom if the mass ratio $M_{GUT}/M_{pl}$ becomes parametrically small. The Planck mass $M_{pl}$ and the mass scale $M_{GUT}$ of the GUT theory are related to the geometry of $B$ and $S$ in the following way:
\begin{equation}
M_{pl}^2\sim \frac{M_s^8}{g_s^2}\mathrm{Vol}(B)\qquad M_{GUT}\sim \mathrm{Vol}(S)^{-\frac{1}{4}}\qquad 1/g^2_\textmd{YM}\sim \frac{M_s^4}{g_s}\mathrm{Vol}(S)\,,
\end{equation}
Therefore one has:
\begin{equation}
\frac{M_{GUT}}{M_{pl}}\sim g^2_\textmd{YM}\frac{\mathrm{Vol}(S)^{3/4}}{\mathrm{Vol}(B)^{1/2}}
\end{equation}
There are two ways to achieve a small value for $M_{GUT}/M_{pl}$, now commonly referred to as the physical and the mathematical decoupling limit. In the physical decoupling limit the volume of the $GUT$ brane $S$ is kept finite while $\mathrm{Vol}(B)\rightarrow\infty$. The mathematical decoupling limit takes $\mathrm{Vol}(S)\rightarrow 0$ for finite volume of $B$. The two limits may not be equivalent in the sense that they may be implemented by tuning different K\"ahler parameters. The volumes of $B$ and $S$ can be determined in terms of the K\"ahler form $J$ of the ambient toric variety restricted to $B$: 
\begin{equation}
\label{volumes}
\mathrm{Vol}(B)=J^3\qquad \mathrm{Vol}(S)=S\cdot J^2
\end{equation}
In order for the volumes to be positive we must find a basis of the K\"ahler cone $K_i$ where, by definition, $J$ can be written as $J=\sum_ir_iK_i$ with $r_i>0$. The existence of mathematical and physical decoupling limits can be deduced from the moduli dependence of these volumes.

Having found a suitable base manifold we can also study matter curves and Yukawa couplings. The curve classes $M$ of the matter curves can be expressed in terms of the toric divisors of the ambient space. The genus of the matter curves can be computed using the first Chern class of the matter curve and the triple intersection numbers:
\begin{equation}
c(M)=\frac{\prod_i(1+D_i)}{(1+B)(1+S)(1+M)}
\end{equation}
Here we have assumed that $M$ is irreducible. 
After expanding this expression to get $c_1(M)$, the Euler number is obtained by the following intersection product:
\begin{equation}
\chi(M)=2-2g(M)=c_1(M)\cdot M\cdot S
\end{equation}
The genus of a matter curve gives us information about the number of moduli the curve has. Since these moduli have to be stabilized, matter curves of low genus are desirable from a phenomenological point of view. In the generic situation the equations specifying the Yukawa points can be expressed as classes $Y_1,Y_2$ in terms of the toric divisors. The number of Yukawa points is then given by the following intersection product:
\begin{equation}
n_{\textrm{Yukawa}}=S\cdot Y_1\cdot Y_2
\end{equation}
In order to account for the Standard Model Yukawa couplings only a small number of Yukawa points is needed. 
\subsubsection*{Fourfold}
Given a base manifold $B$ one can construct a Calabi-Yau fourfold $X_4$ which is an elliptic fibration over $B$. As described in the next section this can be done systematically using toric geometry. However, not all of the desirable features of global F-theory models are automatic in this construction. The main requirement on $X_4$ is that it is a complete intersection of two hypersurfaces. Furthermore these hypersurfaces must have a specific structure (\ref{cicydef}). 
In order to be able to use powerful mathematical tools we furthermore have to require that there exists a nef-partition (cf. section \ref{sec-hypercicy}) which is compatible with the elliptic fibration. When this elementary requirement is satisfied we can engineer a GUT model. This is done in two steps: first we have to identify the GUT divisor $S$, given by the equation $w=0$ in $B$ within the Calabi-Yau fourfold. The second step is to impose the GUT group. This amounts to explicitly imposing the factorization conditions such as (\ref{su5so10tate}) on the Tate model. This means that we have to remove all those monomials in (\ref{tate}) which do not satisfy the factorization constraints. This amounts to fixing a number of complex structure moduli of $X_4$. 

Recently there has been active discussion in the literature how to globally define fluxes in F-theory models \cite{Blumenhagen:2009yv,Marsano:2009wr,Grimm:2009yu,Hayashi:2010zp,Marsano:2010ix,Chung:2010bn,Marsano:2010sq,Collinucci:2010gz,Dolan:2011iu}. In F-theory model building fluxes enter at several crucial points. Gauge flux along the matter curve is needed in order to generate chiral matter. Breaking of the GUT group to the Standard Model gauge group can be achieved by turning on $U(1)$-flux. Furthermore, in $SU(5)$ F-theory GUTs we need global $U(1)$s in order to forbid dimension $4$ proton decay operators. In $SO(10)$ F-theory GUTs they are needed in order to obtain chiral matter \cite{Chen:2010tg,Chen:2010ts}.  A general global description of the fourform flux $G_4$ is still missing. In \cite{Donagi:2009ra} an auxiliary construction involving spectral covers that factorize was used to describe fluxes locally in the vicinity of the GUT brane. It has been shown in \cite{Grimm:2010ez,Marsano:2010ix,Chung:2010bn} that under certain circumstances the information captured by the spectral cover can be encoded in the Tate model, and is therefore global. However, this need not be the case \cite{Hayashi:2010zp}. In \cite{Grimm:2010ez} it has been shown that a spectral cover which factorizes is generically globally defined for ``$U(1)$-restricted Tate models''. This is achieved by imposing a global $U(1)_X$ symmetry in the elliptic fibration. In terms of the Tate model this is achieved by setting $a_6=0$.
\section{Ingredients and Techniques from Toric Geometry}
\label{sec-toric}
In the previous section we have introduced quantities which encode important information about F-theory GUT models in the geometry of the base manifold and the Calabi-Yau fourfold. In this section we will provide the tools to calculate them. The input data needed for these calculations can be obtained by using toric geometry. After giving the basic definitions we will discuss how to describe hypersurfaces and complete intersections of hypersurfaces in toric ambient spaces. Then we explain how to obtain the intersection ring and the K\"ahler cone, or dually, the Mori cone. Finally we will discuss how to use the computer program PALP \cite{Kreuzer:2002uu} for calculations in toric geometry. This discussion of toric geometry has been compiled with a view towards the applications in F-theory model building. It is by no means an exhaustive description of this vast subject which brings together algebraic geometry and combinatorics.
\subsection{Toric Varieties}
\label{sec-toricdef}
We start by defining a toric variety $X$ of dimension $n$ as the following quotient:
\begin{equation}
\label{torvar}
X=(\mathbb{C}^r-Z)/((\mathbb{C}^{\ast})^{r-n}\times G),
\end{equation} 
where $G$ is a finite abelian group, $(\mathbb{C}^{\ast})^{r-n}$ describes the action of an algebraic $(r-n)$-torus and $Z\subset\mathbb{C}^r$ is an exceptional set which tells us which combinations of coordinates are not allowed to vanish simultaneously. The simplest example is $\mathbb{CP}^2$, where the $\mathbb{C}^{\ast}$-action is given by $(z_1,z_2,z_3)\sim (\lambda z_1,\lambda z_2, \lambda z_3)$, $\lambda\in\mathbb{C}^{\ast}$, the exceptional set is $Z=\{z_1=z_2=z_3=0\}$ and $G$ is trivial. Thus, as is well-known: $\mathbb{CP}^2=(\mathbb{C}^3-\{z_1=z_2=z_3=0\})/((z_1,z_2,z_3)\sim (\lambda z_1,\lambda z_2, \lambda z_3))$. 

The crucial fact about toric geometry is that the geometric data of the toric variety can be described in terms of combinatorics of cones and polytopes in dual pairs of integer lattices. The information about the toric variety is encoded in a fan $\Sigma$, which is a collection of strongly convex rational polyhedral cones where all the faces and intersections of pairs of cones also belong to the fan. 'Strongly convex' means that all the cones of the fan have an apex at the origin, 'rational' means that the rays that span the cone go through points of the lattice. We denote by $\Sigma^{(n)}$ the set of all $n$--dimensional cones. In order to define the fan we use the fact that a toric variety $X$ contains an $n$-torus $T=(\mathbb{C}^{\ast})^n$ as a dense open subset whose action extends to $X$. Parametrizing $T$ by coordinates $(t_1,\ldots,t_n)$, one defines the character group $M=\{\chi:T\rightarrow \mathbb{C}^{\ast}\}$ and the one-parameter subgroups $N=\{\lambda:\mathbb{C}^{\ast}\rightarrow T\}$. $M$ and $N$ can be identified with integer lattices that are isomorphic to $\mathbb{Z}^n$. Given a point $m\in M$, the character is given by $\chi^m(t)=t_1^{m_1}\ldots t_n^{m_n}\equiv t^m$. This is a holomorphic function on $T$, and descends to a rational function on the toric variety $X$. For every $u\in N$, $\lambda$ is defined as $\lambda^u(\tau)=(\tau^{u_1},\ldots,\tau^{u_n})$ for $\tau\in\mathbb{C}^{\ast}$. The fan $\Sigma$ and its cones $\sigma$ are defined on the real extension $N_{\mathbb{R}}$ of $N$. The lattices $M,N$ are dual due to the composition $(\chi\circ\lambda)(\tau)=\chi(\lambda(\tau))=\tau^{\langle\chi,\lambda\rangle}$, where $\langle\chi^m,\lambda^u\rangle=m\cdot u$ is the scalar product. 

The $M$-lattice encodes the data about regular monomials in $X$, the $N$-lattice captures the information about the divisors. The divisors defined by $\chi^m=0$ can be decomposed in terms of irreducible divisors $D_j$: $\mathrm{div}(\chi^m)=\sum_{j=1}^ra_jD_j$. These divisors are principal divisors, i.e. divisors of meromorphic function where $D_j$ correspond to poles or zeros and the $a_j$ are orders of the pole/zero. The coefficients $a_j(m)\in\mathbb{Z}$ are unique, and there exists a map $m\rightarrow a_j(m)=\langle m,v_j\rangle$ with $v_j\in N$. Thus there is a vector $v_j$ for every irreducible divisor $D_j$. The $v_j$ are the primitive generators of the one-dimensional cones $\rho_j$ (i.e. rays) in the fan $\Sigma$. The convex hull of the $v_j$ defines a polytope $\Delta^{\ast}=\mathrm{conv}\{v_j\}$. Locally, we can write the divisors as $D_j=\{z_j=0\}$, where $z_j$ is regarded as a local section of a line bundle. $D_j$ are called toric divisors. There are linear relations among the $v_j\in\Delta^{\ast}$ which translate into linear relations among the toric divisors.

In order to make contact with the definition (\ref{torvar}) of $X$, we view the $\{z_j\}$ as global homogeneous coordinates $(z_1:\ldots :z_r)$. If all $z_j$ are non-zero the coordinates $(\lambda^{q_1}z_1:\ldots:\lambda^{q_r}z_r)\sim(z_1:\ldots:z_r)$ with $\lambda\in\mathbb{C}^{\ast}$ describe the same point on the torus $T$, if $\sum q_jv_j=0$ for $v_j\in N$ as above. Since the $v_j$ live in an $n$-dimensional lattice they satisfy $r-n$ linear relations. If the $v_j$ do not span the $N$-lattice there is a finite abelian group $G$ such that $G\simeq N/(\mathrm{span}\{v_1,\ldots,v_r\})$. Identifications coming from the action of $G$ have to be added to the identifications between the homogeneous coordinates coming from the torus action. Having introduced the fan $\Sigma$, we are also able to specify the exceptional set $Z$ that tells us where the homogeneous coordinates are not allowed to vanish simultaneously: a subset of coordinates $z_j$ is allowed to vanish simultaneously if and only if there is a cone $\sigma\in\Sigma$ containing all the corresponding rays $\rho_j$. To be more precise, the exceptional set is the union of sets $Z_I$ with minimal index sets $I$ of rays for which there is no cone that contains them: $Z=\cup_I Z_I$. This is equivalent to the statement that the corresponding divisors $D_j$ intersect in $X$. Putting the pieces together we arrive at the definition (\ref{torvar}).

There are two important properties of the fan $\Sigma$ which translate into crucial properties of the toric variety $X$. Firstly, $X$ is compact if and only if the fan is complete, i.e. if the support of the fan covers the $N$-lattice: $|\Sigma|=\bigcup_{\Sigma}\sigma=N_{\mathbb{R}}$. Secondly, $X$ is non-singular if and only if all cones are simplicial and basic, which means that all cones $\sigma\in\Sigma$ are generated by a subset of a lattice basis of $N$. Singularities can be removed by blow-ups, where singular points are replaced by $\mathbb{P}^{n-1}$s. All the singularities of a toric variety can be resolved by a series of blowups. These correspond to subdivisions of the fan. In order to completely resolve all singularities one must find a maximal triangulation of the fan. In many cases it is sufficient to find a maximal triangulation of the polytope $\Delta^{\ast}$. 

Finally, let us emphasize the significance of the homogeneous weights $q_i$. In general there will be a full $(r-n)\times r$ matrix $Q_{ij}$, called weight matrix, whose $(r-n)$ lines encode the $\mathbb{C}^{\ast}$-actions. Since each of the ${z_j}$ corresponds to an irreducible divisor in $X$, the columns of the weight matrix define a homology basis of the divisors $D_j$. In physics language the weights $q_i$ are the $U(1)$-charges in the gauged linear sigma model that defines the toric variety $X$. Note that the weights contain all the information to recover the  $M$- and $N$-lattice. With the weight matrix as input this can be done using PALP.
\subsection{Hypersurfaces and complete intersections}
\label{sec-hypercicy}
Having defined a toric variety we go on to discuss hypersurfaces and complete intersections of hypersurfaces in toric varieties. The hypersurface equations are sections of non-trivial line bundles. The information of these bundles can be recovered from their transition functions. In this context we introduce the notions of Cartier divisors and Weil divisors. A Cartier divisor is given, by definition, by rational equations $f_{\alpha}=0$ and regular transition functions $f_{\alpha}/f_{\beta}$ on the overlap of two coordinate patches $U_{\alpha},U_{\beta}$. Cartier divisor classes determine the Picard group $\mathrm{Pic}(X)$ of holomorphic line bundles. Weil divisors are finite formal sums of irreducible varieties of codimension $1$. On a toric variety the Chow group $A_{n-1}(X)$ modulo linear equivalence is generated by the $T$-invariant irreducible divisors $D_j$ modulo the principal divisors $\mathrm{div}(\chi^m)$, $m\in M$. A Weil divisor of the form $D=\sum a_jD_j$ is Cartier if there exists an $m_{\sigma}\in M$ for each maximal cone $\sigma\in\Sigma^{(n)}$ such that $\langle m_{\sigma},v_j\rangle=-a_j$ for all rays $\rho_j\in\sigma$. If $X$ is smooth then all Weil divisors are Cartier. If $X$ is compact and $D$ is Cartier then $\mathcal{O}(D)$ is generated by global sections if and only if $\langle m_{\sigma},v_j\rangle>-a_j$ for $\sigma\in\Sigma^{(n)}$ and $\rho_j\not\subset\sigma$. If this is the case for $v\in\sigma$, $\psi_D(v)=\langle m_{\sigma},v\rangle$ is a strongly convex support function. With that we can define a polytope $\Delta_D=\{m\in M_{\mathbb{R}}: \langle m_{\sigma},v_j\rangle\geq -a_j\}$. This is a convex lattice polytope in $M_{\mathbb{R}}$ whose lattice points provide global sections of the line bundle $\mathcal{O}(D)$ corresponding to the divisor $D$. $D$ is generated by global sections if and only if $\Delta_D$ is the convex hull of $\{m_{\sigma}\}$. Furthermore, $D$ is ample if and only if $\Delta_D$ is $n$-dimensional with vertices $m_{\sigma}$ for $\sigma\in\Sigma^{(n)}$ and with $m_{\sigma}\neq m_{\tau}$ for $\sigma\neq\tau\in\Sigma^{(n)}$. Finally $D$ is called base point free if and only if $m_\sigma\in\Delta_D$ for all $\sigma\in\Sigma^{(n)}$. Base point freedom is a sufficient condition for a hypersurface defined by $D$ to be regular: Bertini's theorem states that the singular points of $D$ are the base locus and the singular points inherited from the ambient space. The absence of base points implies that $D$ can be deformed transversally in every point and therefore generically avoids the singularities of the ambient space. Thus, a base point free $D$ is regular. We emphasize however that base point freedom is not a necessary condition for the regularity of $D$.

Equations for hypersurfaces or complete intersections are sections of line bundles $\mathcal{O}(D)$ given by the following Laurent polynomial:
\begin{equation}
f=\sum_{m\in\Delta_D\cap M}c_m\chi^m=\sum_{m\in\Delta_D\cap M}c_m\prod_jz_j^{\langle m,v_j\rangle}
\end{equation}
In an affine patch $U_{\sigma}$ the local section $f_{\sigma}=f/\chi^{m_{\sigma}}$ is a regular function. Given a polytope $\Delta_D\in M$, we can define the polar polytope $\Delta_D^{\circ}$ by $\Delta_D^{\circ}=\{y\in N_{\mathbb{R}}:\:\langle x,y\rangle\geq -1\:\forall x\in\Delta_D\}$. It can be shown \cite{batyrev93} that the Calabi-Yau condition for hypersurfaces requires that $\Delta_D\subseteq M_{\mathbb{R}}$ is polar to $\Delta^{\ast}=\Delta_D^{\circ}\subseteq N_{\mathbb{R}}$, where $\Delta^{\ast}$ is the convex hull of the $v_j\in N$ as defined in section \ref{sec-toricdef}. A lattice polytope whose polar polytope is again a lattice polytope is called reflexive. For reflexive polytopes $(\Delta,\Delta^{\circ})$ there exists a combinatorial formula for the Hodge numbers \cite{batyrev93}:
\begin{equation}
h_{1,1}(X_{\Delta})=h_{\mathrm{dim}\Delta-2,1}(X_{\Delta^{\circ}})=l(\Delta^{\circ})-1-\mathrm{dim}\Delta-\sum_{\mathrm{codim}(\theta^{\circ})=1}l^{\ast}(\theta^{\circ})+\sum_{\mathrm{codim}(\theta^{\circ})=2}l^{\ast}(\theta^{\circ})l^{\ast}(\theta)
\end{equation}
where $\theta$ and $\theta^{\circ}$ is a dual pair of faces of $\Delta$ and $\Delta^{\circ}$. Furthermore, $l(\theta)$ is the number of lattice points of a face $\theta$ and $l^{\ast}(\theta)$ is the number of its interior lattice points. 

In our discussion of F-theory model building we also encounter complete intersection Calabi-Yaus. The concept of polar pairs of reflexive lattice polytopes can be generalized as follows:
\begin{eqnarray}                                        \label{cicy-nef}
        \Delta=\Delta_1+\ldots+\Delta_r&& \Delta^{\circ}
        =\langle \nabla_1,\ldots,\nabla_r\rangle_{\mathrm{conv}}\nonumber\\
                &(\nabla_n,\Delta_m)\geq-\delta_{nm}&\\
        \nabla^{\circ}=\langle \Delta_1,\ldots,\Delta_r\rangle_{\mathrm{conv}}
        &&\nabla=\nabla_1+\ldots+\nabla_r\nonumber
\end{eqnarray}
Here $r$ is the codimension of the Calabi-Yau and the defining equations $f_i=0$ are sections of $\mathcal{O}(\Delta_i)$. The decomposition of the $M$-lattice polytope $\Delta\subset M_{\mathbb{R}}$ into a Minkowski sum\footnote{The Minkowski sum $A+B$ of two sets $A,B$ is defined as follows: $A+B=\{a+b|a\in A, b\in B\}$.} $\Delta=\Delta_1+\ldots+\Delta_r$ is dual to a nef (numerically effective) partition of the vertices of a reflexive polytope $\nabla\subset N_{\mathbb{R}}$ such that the convex hulls $\langle\nabla_i\rangle_{\mathrm{conv}}$ of the respective vertices and $0\in N$ only intersect at the origin. The nef-property means that the restriction of the line bundles associated to the divisors specified by the N-lattice points to any algebraic curve of the variety are non-negative. There exists a combinatorial formula for the Hodge numbers \cite{batyrevborisov} which has been implemented in PALP. 

In many string theory applications, and in particular also in F-theory, the fibration structure of a Calabi-Yau manifold is of great interest. For Calabi-Yaus which can be described in terms of toric geometry the fibration structure can be deduced from the geometry of the lattice polytopes. If we are looking for toric fibrations where the fibers are Calabi-Yaus of lower dimensions, we have to search for reflexive sub-polytopes of $\Delta^{\circ}$ which have appropriate dimension. Given a base $b$ and a fiber $f$, the fibrations descend from toric morphisms of the ambient spaces corresponding to a map $\phi:\Sigma\rightarrow\Sigma_b$ of fans in $N$ and $N_b$, where $\phi:N\rightarrow N_b$ is a lattice homomorphism such that for each cone $\sigma\in\Sigma$ there is a cone $\sigma_b\in\Sigma_b$ that contains the image of $\sigma$. The lattice $N_f$ for the fiber is the kernel of $\phi$ in $N$. The fiber polytope is then defined as follows: $\Delta_f^{\circ}=\Delta^{\circ}\cap N_f$. In order to guarantee the existence of a projection one must find a triangulation of $\Delta_f^{\circ}$ and extend it to a triangulation of $\Delta^{\circ}$. For each choice of triangulation the homogeneous coordinates corresponding to the rays in $\Delta_f^{\circ}$ can be interpreted as coordinates of the fiber. 
\subsection{Intersection ring and Mori cone}
\label{sec-intmori}
Two further pieces of data that are necessary in many string theory calculations are the intersection numbers of the toric divisors and the Mori cone, which is the dual of the K\"ahler cone. Inside the K\"ahler cone the volumes such as (\ref{volumes}) are positive. Thus, in the context of F-theory model building the K\"ahler cone is needed in order to make statements about a decoupling limit. 

Let us start with discussing the intersection ring. For a compact toric variety $X_{\Sigma}$ the intersection ring is of the form $\mathbb{Z}[D_1,\ldots,D_r]/\langle I_{lin}+I_{non-lin}\rangle$. The two ideals to be divided out take into account linear and non-linear relations between the divisors. The linear relations have the form $\sum_j\langle m,v_j\rangle D_j$, where $m\in M$ form a set of basis vectors in the M-lattice. The non-linear relations are denoted by $R=\cup R_I$ where the $R_I$ are of the form $R_I=D_{j_1}\cdot\ldots\cdot D_{j_k}=0$. They come from the exceptional set $Z=\cup Z_I$ defined in section \ref{sec-toricdef}, which determines which homogeneous coordinates are not allowed to vanish at the same time. As mentioned before, this is the case when a collection of rays $\rho_{j_1},\ldots,\rho_{j_k}\in N$ is not contained in a single cone. The non-linear relations $R$ generate the ideal $I_{non-lin}$ which is called Stanley-Reisner ideal. Thus, the intersection ring $A_{\ast}(\Sigma)$ of a non-singular toric variety has the following form:
\begin{equation}
\label{intring}
A_{\ast}(X_{\Sigma})=\mathbb{Z}[D_1,\ldots,D_r]/\langle R,\sum_j\langle m,v_j\rangle D_j\rangle
\end{equation}
The definition of the intersection ring holds for non-singular toric varieties but may be generalized to the case where $X_{\Sigma}$ is simplicial projective. This means that the toric variety may be singular but still all the cones of the fan $\Sigma$ are simplicial. Such a situation may occur for example if we choose a non-maximal triangulation of the polytope $\Delta^{\ast}$. In this case the intersection numbers take values in $\mathbb{Q}$. To compute the Stanley-Reisner ideal in the non-singular case one must find a maximal triangulation of the fan $\Sigma$ or the polytope $\Delta^{\ast}$. In order to get intersection numbers we still have to fix a normalization: for a maximal simplicial cone $\sigma\in\Sigma^{(n)}$ spanned by $v_{j_1},\ldots,v_{j_n}$ we fix the intersection numbers of the corresponding divisors to be $D_{j_1}\cdot\ldots D_{j_n}=1/\mathrm{Vol}(\sigma)$, where $\mathrm{Vol}(\sigma)$ is the lattice volume of $\sigma$ (i.e. the geometric volume divided by the volume $1/n!$ of a basic simplex). If $X$ is non-singular the volume is $1$. Using the intersection ring one can compute the total Chern class of the tangent bundle $T_X$ of $X$ which is given by the following formula: $c(T_X)=\prod_{j=1}^r(1+D_j)$. 

So far, we have only discussed the intersection ring of the toric variety $X$. However in many applications we rather need the intersection numbers for divisors on a hypersurface given by a divisor $D$ in $X$. Here we can make use of the restriction formula that relates the intersection form on the hypersurface divisor to the intersection form on $X$:
\begin{equation}
D_{j_1}\cdot\ldots\cdot D_{j_{n-1}}|_D=D_{j_1}\cdot\ldots\cdot D_{j_{n-1}}\cdot D|_X
\end{equation}
This allows us to compute the intersection ring of $D$ from the intersection ring of $X$. In (\ref{intring}) restriction to $D$ amounts to computing the ideal quotients of $I_{lin}$ and $I_{non-lin}$ with the ideal generated by $D$. By adjunction the Chern class for the hypersurface specified by $D$ is $c(D)=\prod_{j=1}^r(1+D_j)/(1+D)$. 

In order to be able to calculate all the quantities defined in section \ref{sec-fgeom} we miss one more ingredient: the Mori cone. By definition, the Mori cone is the dual of the K\"ahler cone. We need the information about the K\"ahler cone in order to be able to compute the volumes of divisors. By definition the volumes will be positive inside the K\"ahler cone. The Mori cone is generated by $l^{(1)},\ldots,l^{(k)}$, where $k=r-n$ if the fan $\Sigma$ is simplicial. Otherwise the number of Mori generators can be larger. The Mori cone $L$ is then defined as follows: $L=\mathbb{R}_{\geq 0}l^{(1)}+\ldots+\mathbb{R}_{\geq 0}l^{(k)}$. For the calculation of the Mori cone we also require a maximal triangulation of $\Delta^{\ast}$. Given such a triangulation the Mori generators can be determined as follows \cite{Berglund:1995gd}: take every pair of $n$-dimensional simplices $(S_k,S_l)$ which have a common $n-1$-dimensional simplex $s_{kl}=S_k\cap S_l$. Then find the unique linear relation $\sum_il_i^{k,l}v_i=0$ with $v_i\in S_k\cup S_l$ where the $l_i^{k,l}$ are minimal integers and the coefficients of the points in $(S_k\cup S_l)\backslash (S_k\cap S_l)$ are non-negative. The Mori generators are then the minimal integers $l^{(a)}$ by which every $l^{k,l}$ can be expressed as positive integer linear combinations. There is an equivalent algorithm to determine the Mori generators due to Oda and Park \cite{OdaPark} which has been implemented in an unreleased version of PALP \cite{maxnils}. Note that the relations $\sum_{i=1}^rl_i^{(a)}D_i=0$ define the ideal $I_{lin}$ in (\ref{intring}). Assembling the Mori vectors into a $k\times r$-matrix, the columns of the matrix encode inequalities for the values of the K\"ahler parameters. Solving these inequalities yields a basis $K_i$ of the K\"ahler cone such that the K\"ahler form of $X$ can be written as $J=\sum_i r_iK_i$ with $r_i>0$. Note that this prescription computes the K\"ahler cone of the toric variety $X$. It is often assumed that this is a good approximation for the K\"ahler cone of a hypersurface in $X$.     
\subsection{Toric Calculations using PALP and other Software}
In string theory and F-theory we deal with compactifications on Calabi-Yau threefolds and fourfolds. In F-theory model building the base manifold $B$ is a hypersurface in a four-dimensional toric ambient space. The fourfolds are complete intersections in a six-dimensional toric space. The associated lattice polytopes live in four- and six-dimensional integer lattices and typically have a large number of points. It is in general not possible to do calculations without computer support. There exist several software packages which are useful for particular aspects in toric geometry. In this section we will mostly focus on the program PALP \cite{Kreuzer:2002uu}. Before that, let us mention some other useful programs: Schubert by Katz and Str\o mme is a Maple package for calculations in intersection theory. TOPCOM \cite{Rambau:TOPCOM-ICMS:2002} computes triangulations of point configurations. Singular \cite{DGPS} is a powerful computer algebra program which is optimized for calculations with polynomial rings, such as the intersection ring. A recent addition is cohomCalg \cite{Blumenhagen:2010pv} which can compute line bundle-valued cohomology classes over toric varieties.

Let us now discuss some features and applications of PALP \cite{Kreuzer:2002uu}, which stands for ``Package for Analyzing Lattice Polytopes''. It consists of several programs. 
\begin{itemize}
\item {\tt poly.x} computes the data of a lattice polytope and its dual if the polytope is reflexive. The input can be either a weight matrix or the points of a polytope in the M-lattice or the N-lattice. Apart from the polytope data {\tt poly.x} computes Hodge numbers of the associated Calabi-Yau hypersurfaces, information about fibrations and other data. {\tt poly.x} has been extended with several features that include information about the facets of the polytope, data of Fano varieties and conifold Calabi-Yaus. In \cite{Batyrev:2008rp,Kreuzer:2009is} this extension of PALP has been used to find new Calabi-Yau manifolds with small $h^{1,1}$ which are obtained from known Calabi-Yau threefolds via conifold transitions. The full set of options in PALP can be obtained with {\tt poly.x -h} and {\tt poly.x -x} for extended options. 
\item The program {\tt nef.x} can be used for complete intersection Calabi-Yaus. It takes the same input as {\tt poly.x} and computes the polytope data, nef partitions and Hodge numbers as well as information about fibrations. There are several extended options which include most notably the data of the Gorenstein cones (cf. \cite{Batyrev:1994ac} for the definition and construction in toric geometry) in the $M/N$-lattice. 
\item {\tt cws.x} creates weight systems and combined weight systems of polytopes of dimension to be specified in the input. 
\item {\tt class.x} classifies reflexive polytopes by searching for sub-polytopes of a Newton polytope associated to a combined weight system. 
\end{itemize}
Apart from recent applications in F-theory model building, which we will discuss in the next section, PALP has been used in many other contexts. A data base of Calabi-Yau threefolds has been generated by listing all $473\,800\,776$ reflexive polyhedra in four dimensions \cite{Kreuzer:2000xy}. In view of the landscape problem in string theory the statistics of the polytope data is also of interest \cite{Kreuzer:2008nu}. Some of the most recent extensions of PALP which we will mention below have already been used in \cite{Collinucci:2008sq,Chen:2010ts,Knapp:2011wk}.
\subsection{Application to F-theory GUTs}
In this section we make the connection to F-theory model building and discuss how the calculations discussed in section \ref{sec-fgeom} can be carried out explicitly. The approach discussed here is used in \cite{Chen:2010ts,Knapp:2011wk}. Our aim is a systematic construction of a large class of examples of global F-theory models. The first step is the construction of the base manifold $B$. In \cite{Chen:2010ts} we have obtained a set of geometries by systematically constructing weight matrices associated to point and curve blowups on Fano hypersurfaces in $\mathbb{P}_4$. In \cite{Knapp:2011wk} we have considerably extended this class of models by defining hypersurfaces in a subset of the toric ambient spaces described by the $473\,800\,776$ reflexive polyhedra in four dimensions \cite{Kreuzer:2000xy}. Concretely, we have restricted ourselves to configurations where the N-lattice polytopes have at most nine points. As one can check for example at \cite{database}, there are $1088$ such polytopes. We used PALP to recover the toric data of the ambient space and considered all non-negatively curved hypersurfaces in these ambient spaces. In order to be able to perform the calculations outlined in section \ref{sec-fgeom} we must compute the intersection ring and the Mori cone. We have achieved this by using an extended version of {\tt poly.x} \cite{maxnils}. The following additional features have been implemented: processing of non-Calabi-Yau hypersurfaces by specifying the hypersurface degrees as input parameters, a calculation of the maximal triangulations of the $N$-lattice polytope, calculation of the Mori cone and the Stanley-Reisner ideal, and calculation of the intersection ring with the help of Singular. Using this data we can identify del Pezzo divisors, check the existence of a decoupling limit and compute the topological properties of matter curves and Yukawa points. In \cite{Knapp:2011wk} we have analyzed a total number of $569\:674$ base manifolds. The resulting geometries are available at \cite{data}.

The next step in the calculation is to construct the Calabi-Yau fourfold $X_4$ which is an elliptic fibration over the base $B$. The toric data of $X_4$ is obtained by extending the weight matrix of $B$. Schematically, this looks as follows:
\begin{equation}
\begin{array}{cccccc}
3&2&1&0&\cdots&0\\
\ast&\ast&0&w_{11}&\cdots&w_{1n}\\
\ast&\ast&0&\cdots&\cdots&\cdots\\
\ast&\ast&0&w_{m1}&\cdots&w_{mn}\\
\end{array}
\end{equation}
Here the $w_{ij}$ denote the entries of the weight matrix associated to $B$. The $\ast$-entries in the extended weight matrix have to be chosen in such a way that the fiber coordinates $x,y$ are sections of $K_B^{-2}$ and $K_B^{-3}$, respectively. These entries of the fourfold weight matrix contain the information about the hypersurface degrees of the base. Not every extended weight system will lead to a Calabi-Yau fourfold of the form (\ref{cicydef}). The calculations can be done using {\tt nef.x}. Several problems can appear: first, there may be no nef partition and therefore our methods do not work. A second conceptual problem is that the polytope corresponding to the extended weight system is not always reflexive. Many of the combinatorial tools used in PALP are only valid for reflexive polytopes. Even though one might have a perfectly fine Calabi-Yau fourfold we cannot apply our technology to them. The third issue is of a technical nature: due to the complexity of the fourfold polytopes one may reach the software bounds of PALP which results in numerical overflows. For the $569\,674$ extended weight systems discussed in \cite{Knapp:2011wk} we find only $27\,345$ reflexive fourfold polytopes which have at least one nef partition. Furthermore there are $18\,632$ reflexive polytopes without a nef partition, $381\,232$ non-reflexive polytopes and $142\,470$ cases with numerical overflow. 

Having found a reflexive fourfold polytope with at least one nef partition is not enough to have a good global F-theory model. If we further demand that the base $B$ has at least one del Pezzo divisor with a mathematical or physical decoupling limit the number of fourfolds decreases significantly. In addition we should also impose some constraints on the regularity of the base. Demanding that $B$ is Cartier leaves us with $16\,011$ good models. Imposing the stronger criterion of base point freedom we are down to $7386$ models. Focusing on these $7\,386$ good geometries we apply the constraint that the nef partition should be compatible with the elliptic fibration. This information can be extracted from the output of {\tt nef.x}. This further reduces the number of geometries to $3978$. 

Having found a good Calabi-Yau fourfold, we can construct a GUT model on every (del Pezzo) divisor. A toric description on how to impose a specific GUT group on a Tate model has been given in \cite{Blumenhagen:2009yv}. The Tate form~(\ref{tate}) implies that the $a_n$ appear in the monomials which contain $z^n$. We can isolate these monomials by identifying the vertex $\nu_z$ in $(\nabla_1,\nabla_2)$ that corresponds to the $z$-coordinate in the Tate model.  All the monomials that contain $z^r$ are then in the following set:
\begin{equation}
A_r=\{w_k\in\Delta_m\::\:\langle\nu_z,w_k\rangle+1=r\}\qquad \nu_z\in\nabla_m,
\end{equation}
where $\Delta_m$ is the dual of $\nabla_m$, which denotes the polytope containing the $z$-vertex. The polynomials $a_r$ are then given by the following expressions:
\begin{equation}
a_r=\sum_{w_k\in A_r}c_k^{m}\prod_{n=1}^2\prod_{\nu_i\in\nabla_n}y_i^{\langle \nu_i,w_k\rangle+\delta_{mn}}\vert_{x=y=z=1}
\end{equation}
Now we can remove all the monomials in $a_r$ which do not satisfy the factorization constraints (\ref{su5so10tate}) of the Tate algorithm. In order to perform this calculation we have to identify the fiber coordinates $(x,y,z)$ and the GUT coordinate $w$ within the weight matrix of the fourfold. We have applied this procedure to every del Pezzo divisor in the $3978$ ``good'' fourfold geometries. Note that the procedure described above can destroy the reflexivity of the polytope, which happens in about $30\%$ of the examples. For $SU(5)$-models we found $11\,275$ distinct models\footnote{Since the procedure has been applied to all del Pezzo divisors in a given base geometry not all these models may have a decoupling limit.} with reflexive polyhedra, for $SO(10)$ GUTs there are $10\,832$. $U(1)$-restricted GUT models \cite{Grimm:2010ez} can be engineered along the same lines. It turns out that $U(1)$-restriction does not put any further constraints on the reflexivity of the polytope.
\section{Outlook}
In this article we have discussed how toric geometry can be used to construct a large number of geometries that can support global F-theory GUTs. Using this technology we could show that elementary consistency constraints greatly reduce the number of possible models. However, due to computational constraints, we did not quite succeed in systematically listing all possible F-theory models within a class of geometries. Such an endeavor would require substantial changes in the computer programs we are using. It is actually quite remarkable that we could make use of PALP for Calabi-Yau fourfolds and non-Calabi-Yau threefolds, since this goes beyond what it was originally designed for.

Let us present a list of suggestions to extend PALP in order to improve the applicability to the current problems in mathematics and physics and to make it more accessible for users. The original purpose of PALP was to solve a classification problem for polytopes. Over the years it has been adjusted and extended in order to be applied to specific problems. Many of the basic routines that were implemented to tackle some special questions could be used in much more general contexts but cannot be easily accessed. Therefore a better modularization of the software is necessary in order to have flexible access to these basic routines. Another problem of PALP is that one has to specify several parameters and bounds such as the number of points in a polytope in a given dimension at the compilation of the program. It would be practical to have fully dynamical dimensions in order to work with a precision tailored to the problem at hand without recompiling.

A fundamental change would be to step away from the description of polytopes and instead use the ray representation which has the full data of the cones. This is necessary if one wants to deal with non-reflexive polytopes. A further extension which has already been partially implemented is to include triangulations, intersection rings and even the calculation of Picard-Fuchs operators needed for mirror symmetry calculations into PALP. The ultimate goal is to have an efficient and versatile program which can be used for toric calculations of all kinds without having to rely on commercial software. Finally a detailed documentation of all the features of PALP would be helpful \cite{andnils}.

As for the search for F-theory models, an extended version of PALP would hopefully help to overcome the problems of non-reflexivity and overflows we have encountered in \cite{Knapp:2011wk}. Apart from finding new examples for physics applications one might also attempt a partial classification of Calabi-Yau fourfolds. Enumerating all toric Calabi-Yau fourfolds may be out of reach or even impossible but for finding all models of type (\ref{cicydef}) one can at least give a prescription for the construction: take each of the $473\,800\,776$ reflexive polyhedra in four dimensions and put in all non-negatively curved hypersurfaces that are not Calabi-Yau. Then construct fourfolds which are elliptic fibrations over these base manifolds. A rough estimate shows that this procedure would yield $\mathcal{O}(10^{11})$ fourfold geometries.\\\\
{\bf Acknowledgments:} In October 2010 Maximilian Kreuzer asked me to be the co-author of this review article. Sadly, he passed away on November 26, 2010 when this work was still in the early stages. I am grateful for many years of collaboration with Max, as well as for his constant support and encouragement. 

I would like to thank my collaborators Ching-Ming Chen, Christoph Mayrhofer and Nils-Ole Walliser for a pleasant and fruitful collaboration on the projects \cite{Chen:2010ts,Knapp:2011wk} this article is based on. Furthermore I thank Emanuel Scheidegger for valuable comments on the manuscript. This work has been supported by World Premier International Research Center Initiative (WPI Initiative), MEXT, Japan.

\addcontentsline{toc}{section}{References}

\providecommand{\href}[2]{#2}\begingroup\raggedright\endgroup

\end{document}